\documentstyle[12pt,openbib]{article}
\begin{document}
\title{\bf 
QCD phase transitions from relativistic hadron models.
}
\vskip .3cm
\author{A. Delfino $^{1}$ \thanks{Partially supported by CNPq of Brasil}, 
Jishnu Dey $^{1,2,3}$ \thanks{Supported by CAPES and FAPESP of
Brasil, work supported in part by DST grant no. SP/S2/K04/93, Govt.
of India, permanent address : 1/10 Prince Golam Md. Road, Calcutta
700 026, India, email : jdey@ift.uesp.ansp.br, jdey@ift.unesp.br} , 
Mira Dey $^{2}$ \thanks{Supported by CAPES of Brasil, work supported
in part by DST grant no. SP/S2/K04/93, Govt. of India, on leave from Lady
Brabourne College, Calcutta 700 017, India, email : mdey@ift.uesp.ansp.br,
mdey@ift.unesp.br}
\, and M. Malheiro $^{1}$ \thanks{Partially supported by CNPq and CAPES of Brasil, Present address: Department of 
Physics, University of Maryland, College Park, Maryland 20742-4111, USA},  
\\ $^{1}$ Instituto de F\'\i sica,
Universidade Federal Fluminense, \\ 24210-340, Niter\'oi,
R. J., Brasil 
\\${^2}$ Instituto de F\'\i sica Te\'orica,
Universidade Estadual Paulista, \\ 01405-900, S\~ao Paulo,
S. P., Brasil 
\\$^{3}$ Dept. de F\'isica, Instituto Tecnol\'ogico da Aeron\'autica, 
\\ CTA, 12228-900, S\~ao Jos\'e Dos Campos, Brasil}
\vspace{.1 cm}
\date{\today }
\maketitle
{\it Abstract} :
The models of translationally invariant infinite nuclear matter in the
relativistic mean field models are very interesting and simple, since the
nucleon can connect only to a constant vector and scalar meson field.  Can
one connect these to the complicated phase transitions of QCD ? For an
affirmative answer to this question, one must consider models where the
coupling constants to the scalar and vector fields must depend on density in
a non-linear way, since as such the models are not explicitly chirally
invariant. Once this is ensured, indeed one can derive a quark condensate
indirectly from the energy density of nuclear matter which goes to zero at
large density and temperature. The change to zero condensate indicates a
smooth phase transition.

\newpage
{\bf 1. Introduction}
\vspace{.5 cm}

There is lot of interest in deriving QCD parameters from various relativistic
hadronic model calculations \cite{cfg}, \cite{cfg2}, \cite{li}. In the
present paper :

(1) we highlight the fact that different hadronic models, - which fit the
binding energy ${\cal E}/ \rho - M$  at the density of the infinite nuclear
matter $\rho _o$, - also gives us the same quark condensate at $\rho _o$. 
But the expected behaviour,  of ${ \langle \bar q q\rangle}\rightarrow 0$ at
high $\rho$ and T, is {\it not built  in the models}. Indeed since the
models are not explicitly chirally invariant, one can combine the behaviour
${ \langle \bar q q\rangle}\rightarrow 0$ with an effective nucleon mass which
is not changing very much at high density. In the next section we will
discuss this in more detail with respect to the recent queries \cite{eric}
regarding chiral symmetry restoration. 

(2) qualitative arguments are given to relate the behaviour of quark condensate
${\langle \bar q q\rangle}$, to the incompressibility of nuclear matter in 
hadronic models (HM, in short).

(3) we stress that to link QCD to hadronic models is important.  The current
attempts to derive QCD parameters from HM may provide important clues that
will help in derivations of NN force from QCD. In QCD there is only one
parameter $\Lambda _{QCD}$ in terms of which one gets the strong coupling
$\alpha _s$. This in turn must control the quark and gluon condensates,
although to our knowledge can only cite heuristic arguments to show this, for
example in ref. \cite{bailin} :

\begin{equation}
 m_{dyn} = 3 e ^{1/6} \Lambda _{QCD}= 300 \; MeV
\label{eq:bailin1}
\end{equation}
where $m_{dyn}$ is the consituent quark mass and 
\begin{equation}
{\frac {4}{3}} \pi \alpha _s{ \langle \bar q q\rangle}_0   = - (m_{dyn})^3
\label{eq:bailin2}
\end{equation}
with $ \Lambda _{QCD} = 130$ MeV.

In nuclear matter, there are more parameters, the binding energy -15.75, the
equilibrium density and the incompressibility which are now leading to a
specific value of ${\langle \bar q q\rangle}_{\rho}$, replacing them, in a
sense, in terms of one parameter. The hadronic models which we consider,
namely the Walecka model \cite{wa} and its modifications \cite{zm}, show
saturation of infinite nuclear matter through large cancellations between the
attractive scalar and repulsive vector potentials. However this underlying NN
force is as yet unknown from the QCD point of view, and in some way the QCD -
construction of the NN force may be helped by the connection between
${\langle \bar q q\rangle}$ and the incompressibility.

The plan of the paper is as follows : in section 2 we discuss the quark
condensate, how it is supposed to relate the scaling of the nucleon effective
mass in a medium, the Hellmann-Feynman way to evaluate it from the quark
condensate. In section 3 the nuclear models are discussed. Section 4 is
devoted to results and discussions at finite density while section 5
discusses the nuclear equation of state. Section 6 contains the finite T
results for zero baryonic density but non-zero scalar density, which we
believe is and will be probed in mid-rapidity reactions with very heavy ions.
Section 7 contains the summary and conclusions. 

\vspace{.5 cm}
{\bf 2. The quark condensate}
\vspace{.5 cm}

The quark condensate in QCD is :
\begin{equation}
    { \langle \bar q q\rangle}_0   = - (230 \pm 30 \;\;MeV)^3
\label{eq:qqbar}
\end{equation}
The non-zero value is due to the breaking of approximate chiral
symmetry by the vacuum, otherwise enjoyed by the QCD Hamiltonian, by
virtue of the smallness of the quark mass $m_q$ :
\begin{equation}
H_{QCD} = H_0 + 2 m_q \bar q q
\label{eq:hqcd}
\end{equation}
the major part of the above being the chirally symmetric $H_0$.
 
A necessary but not sufficient condition for chiral symmetry restoration in
QCD is that the ${ \langle \bar q q\rangle}\rightarrow 0$ at high $\rho$ and
T \cite{eric}. If the symmetry is still not restored the pseudo-Goldstone
particle, the pion, will be massless and one can have pion condensation
\cite{cb}.  Stretching the point further, can one presumably say that the
effective nucleon will also be very light, if the pion becomes light ?
Keeping these in mind we study the density and temperature dependence of the
condensate and the corresponding sigma term. Let us begin with medium effects.
Lattice studies cannot cope with finite density problems yet and QCD sumrule
determination of the nucleon sigma term is not satisfactory- according to
ref.\cite{gc}. In this work  hadronic models of nuclear matter are employed
for our purpose.

In the linear Walecka model \cite{wa}, the condensate goes down linearly upto
1.5 times the nuclear matter density but then tends to {\it increase},
contrary to expectations based on ideas of chiral symmetry restoration. This
is the same kind of result obtained by Li and Ko \cite{li} very recently from
Bonn potential in the relativistic Dirac-Brueckner approach. They also find
the results unpalatable and conclude that effects like dependence of meson
-nucleon coupling constants on the current quark mass and chiral invariance
may become crucial in obtaining a reliable result for the density dependence
of the quark condensate. Indeed we find the contrary result for variants of
the Walecka model given by Zimanyi and Moszkowski (see \cite{zm}), the models
are labelled by ZM and ZM3 respectively where mesons interact non-linearly
and couplings are $\rho$-dependent. In ZM3, the effective nucleon
remains massive but ${\langle \bar q q\rangle}$ goes to zero at high $\rho$.
 
Before details of our calculations we discuss a few relevant points.  In the
QCD sum rule approach, studied by Ioffe \cite{io} and others \cite{rry} one
gets a simple approximate expression for the nucleon mass :
\begin{equation}
M_N = - (8 \pi ^2/M^2) {\langle  \bar q q \rangle}_0
\label{eq:938}
\end{equation}
and the formula is to be evaluated for $M^2 \sim {M_N}^2$.  This shows that
to this order of approximation ( about 10 per cent or so) the nucleon mass is
controlled by the quantity in eq.(\ref {eq:qqbar}).  From this rough analysis
one would conclude that at finite temperature $T$ or density $\rho$ one
should have a scaling
\begin{equation}
\frac {{\langle \bar q q\rangle}_{\rho\; or \; T}} {{\langle
\bar q q \rangle}_0} = [{M_N}^*/M_N]^3
\label{eq:scaling3}
\end{equation}
where ${M_N}^*$ is the effective nucleon mass in finite $\rho$ or $T$. Indeed
this was the idea put forward in \cite{br}. Later it was suggested \cite{cfg}
that more careful analysis of the sum rule yields a linear scaling :
\begin{equation}
\frac {{\langle \bar q q \rangle}_{\rho \; or \; T}} {{\langle \bar q
q \rangle}_0} = [{M_N}{^*}/M_N]
\label{eq:scaling1}
\end{equation}
at least for finite density. In Figs.1a and 1b neither eq.(\ref{eq:scaling3})
nor eq.(\ref{eq:scaling1}) is strictly valid beyond normal nuclear matter
density $\rho _0$ and the cubic scaling is roughly obeyed by the ZM model
(Fig. 1a) while the linear one is preferred by the ZM3 model (Fig.1b). We now
review how these figures can be obtained from standard nuclear matter
calculations.
 
A concrete way of getting the quark condensate was laid down by \cite{cfg2}
using the Hellmann-Feynman theorem :
\begin{equation}
 {\langle \psi (m_q)| {\frac{d}{dm_q}} H_{QCD} | \psi (m_q) \rangle}
  = {\frac{d}{dm_q}} {\langle \psi (m_q)| H_{QCD} | \psi(m_q) \rangle}
\label{eq:hellmann}
\end{equation}
which on using eq.(\ref{eq:hqcd}) yields
\begin{equation}
2 m_q({\langle \bar q q \rangle}_\rho - {\langle \bar q q\rangle}_0)
= m_q {\frac {d{\cal E}}{d m_q}}.
\label{eq:cfg}
\end{equation}
where the subscript $\rho $ and $0$ indicate expectation value of the
relevant operator for the nucleon in nuclear matter with uniform density
$\rho$ and for the vacuum respectively. The expression is the same for finite
$T$. $\cal E$ is the appropriate energy density.  The leading term of the
expression $d{\cal E}/{d m_q}$ of eq.(\ref{eq:cfg}) is the experimentally
known quantity $\sigma _N$
\cite{cfg2} :
\begin{equation}
2 m_q\int d^3x({\langle N|\bar q q|N\rangle} - {\langle \bar q
q\rangle}_0) = \sigma_N = m_q {\frac {d{M_N}}{d m_q}}.
\label{eq:sigma}
\end{equation}
where $|N\rangle $ is the free nucleon at rest. In nuclear matter of volume V
with $A = \rho V$ nucleons, translational invariance makes the quark
condensate density constant and the integral in eq.(\ref{eq:sigma}) gives :
\begin{equation}
 \sigma_A = 2 m_q V({\langle \bar q q\rangle}_\rho - {\langle \bar q
q\rangle}_0).
\label{eq:sigmaA}
\end{equation}
Comparing eq.(\ref{eq:sigmaA}) to eq.(\ref{eq:cfg}) :
\begin{equation}
{\frac {\rho}{A}} \sigma_A = m_q {\frac {d{\cal E}}{d m_q}}
\label{eq:calE}
\end{equation}
and (using the relation of Gell-Mann, Oakes and Renner) :
\begin{equation}
\frac{{\langle \bar q q\rangle}_\rho}{\langle \bar q
q\rangle}_0 = 1 - {\frac {\rho}{A}} \frac{\sigma_A}{{m_\pi }^2{f_\pi
}^2} = 1 - \rho \frac{\sigma_{eff}}{{m_\pi }^2{f_\pi
}^2}
\label{eq:fpimpi}
\end{equation}
with $f_\pi = 93$ MeV and
\begin{equation}
\sigma _{eff} = \frac{\sigma _A}{A}\,=\, m_q {\frac {d({\cal E}/ \rho)}{d
m_q}}\,=\, \sigma _N {\frac {d({\cal E}/ \rho)}{d
M_N}}
\label{eq:sigmaeff}
\end{equation}
is the effective $\sigma $-commutator for a nucleon in the nuclear medium.
If we neglect
$\delta \cal E $, the nucleon kinetic and interaction energy density in
\begin{equation}
{\cal E} = \rho M_N  + \delta {\cal E}
\label{eq:deltaE}
\end{equation}
we get $\sigma _{eff} = \sigma_N$.

Combining the eqns.(\ref {eq:cfg}, \ref {eq:sigmaA} and \ref{eq:deltaE}) we
separate the static nucleon part and the effect of $\delta {\cal E}$ on
$\sigma _{eff}$ is : 
\begin{equation}
\sigma _{eff} =\sigma_N - \frac {{\langle \bar q q \rangle}_{\rho}
^{\delta {\cal E}}} {{\langle \bar q q \rangle}_0}
\frac {{m_ \pi ^2}{f_\pi^2}}{\rho}.
\label{eq:exp}
\end{equation}

\vspace{.5 cm}
{\bf 3. Relativistic nuclear matter models}
\vspace{.5 cm}

To consider in-medium quark condensate, one needs relativistic nuclear matter
\cite{cfg2}. Serot and Walecka \cite{wa} advocated  a relativistic field
theoretic description of nuclear matter based on the nucleon interacting with
the scalar $\sigma$ and the vector $\omega$ meson fields linearly. And this
is one of the models used in ref.\cite{cfg2} to analyze ${\langle
\bar q q\rangle}_\rho$ for $\rho\sim 1.5 \; \rho_o$ at $T=0$. Note, however,
that our results do not exactly match with theirs, since they use different
values for the nuclear binding( -16 MeV and $\rho _0= 0.17 fm^{-3}$), 
but are in qualitative agreement.

Spatially uniform nuclear matter, in spite of its simplicity, admits the
following vital questions :
 
(1) What happens to ${{\langle \bar q q\rangle}_\rho}$ when the $\sigma$
and $\omega$ fields are non-linear, inducing the couplings to be density
dependent? These features were found to be very important say, in reducing
the nuclear matter compressibility.
 
(2) What is the effect on the effective nucleon mass? Is it similar to
that of ${{\langle \bar q q\rangle}_\rho}$ ? What happens at $T\ne 0$ ?
 
Keeping these in mind we choose, apart from Walecka, two other models used in
\cite{zm}. So the models are :

(1) Linear Walecka model where the coupling constants of the nucleon to
$\sigma$ and $\omega$ fields, $g_{\sigma}$ and $g_{\omega}$ respectively,
remain constant with $\rho$.
 
(2) ZM : (the usual one in the literature): It is constructed by changing the
covariant derivative term in the Walecka model in such a way that, after an
appropriate rescaling, the Lagrangian describes the motion of a baryon with
an effective mass $M^{\ast}=m^{\ast}M$ instead of the bare mass M.  This
information goes to the meson-baryon coupling, modifying it to an effective
scalar coupling constant, dependent on $\rho$, while the vector coupling
constant remains the same.
 
(3) ZM3 : A variant of ZM is obtained if, instead of modifying the covariant
derivative term, one simply modifies the kinetic energy term of the baryon.
As before, after an appropriate rescaling, the Lagrangian describes a baryon
of mass $M^*$. This information manifests not only in the scalar-baryon
coupling but also in the vector-baryon coupling. Both of them now depend on
density. The vector and the scalar fields are now coupled.
 
Connected in this way, $$ {\cal L}_{ZM} \equiv {\cal L}_{Walecka}(g_{\sigma}
\rightarrow g_{\sigma} ^{\ast})\,\,and\,\, {\cal L}_{ZM3} \equiv {\cal
L}_{Walecka} (g_{\sigma}\rightarrow g_{\sigma} ^{\ast}\:;\:
g_{\omega}\rightarrow g_{\omega}^{\ast})$$.

The effective coupling constants are given by : \, $g_{\sigma}^{\ast}/
g_{\sigma} = m^{\ast}$\,\,\, and $g_{\omega}^{\ast}/ g_{\omega} =
m^{\ast}$\,\,\,where $\,{M_N^*}/{M_N}\,= \,m^{\ast}(\sigma)\, =\,
\,(1 + g_{\sigma}{\sigma}/M)^{-1}$. The expression for the energy density at
a given temperature T can be found in the mean field approach (MFA),
\begin{equation}
 {\cal E} =
\frac{{g_{\omega}^{\ast}}^{2}}{2m^{2}_\omega} \rho_{b}^2+
\frac{m^{2}_\sigma}{2{g_{\sigma}^{\ast}}^{2}}(M_N-M_N^{\ast})^2+
\frac{\gamma}{(2\pi)^{3}}\int d^{3}k\,E^{\ast}(k)(n_{k} - \bar n_{k}) ,
\label{eq:energy}
\end{equation}
where
\begin{equation}
\rho_{b} = \frac{\gamma}{(2\pi)^{3}}\int d^{3}k\,(n_{k} - \bar n_{k}).
\label{eq:rhob}
\end{equation}
Here the degeneracy factor  $\gamma = 4 $, $n_{k}$ and $\bar n_{k}$
stand for the Fermi-Dirac distribution for baryons and antibaryons
with arguments $ (E^{\ast} \mp \nu)/T $ respectively. The energy $
E^{\ast}(k) = ( k^2 + M_N^{\ast 2} )^{\frac{1}{2}}$ and $\nu$ is an effective
chemical potential which preserves the number of baryons and antibaryons in
the ensemble.  The effective nucleon mass $M_N^*$, is obtained from the
minimization of $\,{\cal E}\,$. From eq.(\ref{eq:cfg})
\begin{equation}
\frac {{\langle \bar q q \rangle}_{\rho \,,\, T}} {{\langle \bar q
q \rangle}_0} = 1 - {\frac {m_q}{{m_\pi }^2{f_\pi}^2}}[\frac{\partial
{\cal E}}{\partial M_N} \frac{\partial{M_N}}{\partial m_q} +
\frac {\partial {\cal E}}{\partial m_{\omega}} \frac {\partial m_{\omega}}
{\partial m_q} + \frac{\partial{\cal E}}{\partial m_{\sigma}}
\frac{\partial m_{\sigma}}{\partial m_q} + \frac{\partial {\cal E}}
{\partial g_{\omega}} \frac{\partial g_{\omega}} {\partial m_q}
 + \frac{\partial {\cal E}}{\partial g_{\sigma}} \frac{\partial
g_{\sigma}} {\partial m_q}] .
\label{eq:qqrhoT}
\end{equation}
We adopt the rules \cite{cfg2},
\begin{equation}
\frac{\partial m_{\sigma}}{\partial m_q} =
\frac{m_\sigma}{M_N} {\frac {\sigma_N}{m_q}}, \;\; and \;\;
\frac {\partial m_{\omega}}{\partial m_q} =
\frac {m_\omega}{M_N}{\frac{\sigma_N}{m_q}},
\label{eq:Mmeson}
\end{equation}
and for the Walecka model the variation of the meson-couplings with $m_q$ is
unspecified and neglected. Calculating the derivatives in eq.(\ref
{eq:qqrhoT}), by using eq.(\ref {eq:energy}), we obtain a unified expression
for $ {\langle \bar q q \rangle}_{\rho}$ at some T:
\begin{equation}
\frac {{\langle \bar q q \rangle}_{\rho \, ,\, T}} {{\langle \bar q
q \rangle}_0} = 1 - {\frac {\sigma_N}{{m_\pi }^2{f_\pi}^2}}
[\frac{{m_\sigma}^2} {{g^*_\sigma}^2} (M_N - M^*_N) + (1 + \alpha
)\frac{{m_\sigma}^2} {{g^*_\sigma}^2 M_N}(M_N - M^*_N)^2 - (1 +
\beta)\frac{{g_{\omega}^{\ast}}^{2}}{m^{2}_\omega M_N} \rho_{b}^2]  ,
\label{eq:qqrho}
\end{equation}
where, in terms of $\alpha$ and $\beta$, the models are describes as :
Walecka\,($\alpha= \beta = 0$),\,ZM\,($\alpha=\,1$\,and\,$\beta \,= \,0$) and
ZM3\,($\alpha = \beta\,= 1$\,). The terms multiplied by $\alpha$ and $\beta$
are obtained from the variation of the scalar and vector effective couplings
with $m_q$ respectively.

\vspace{.5 cm}
{\bf 4. Results and Discussion for finite $\rho $}
\vspace{.5 cm}

The parameters of the models are presented in Table I to fix the binding
energy per particle to be -15.75  MeV at $\rho _0 = 0.15  fm^{-3}$.
It is found that the $ {{\langle \bar q q \rangle}_{\rho}}/ {{\langle \bar q q
\rangle}_0}$ and  $\sigma_{eff}$ are almost model independent upto a density
$\rho _0 $ (Fig. 2). Throughout this paper $\sigma_N$ = 45 MeV. With this, at
$\rho =\rho _0$, $\sigma _{eff}$ = 44.245 MeV for all the models.  As
expected \cite{cfg2}, the reduction is less than 2$\%$. However, as evident
in Fig.3, for high densities this reduction, $10 - 20 \% $ for ZM models, is
not neglegible anymore.  We stress here that ${{\langle \bar q
q\rangle}_{\rho _o}}/ {{\langle \bar q q \rangle}_0} = 0.69 $, for all the
models.  We calculate the contribution of $\delta {\cal E}$, using
eq.(\ref{eq:exp}), to the condensate and find it negligible at small
densities. For example at $\rho=\rho _o$, ${{{\langle \bar qq\rangle
}_{\rho}^{\delta {\cal E}}}/ {{\langle \bar q q \rangle}_0}} \simeq 0.005$ as
compared to 0.69 for the total. Further, ${{\langle \bar q q
\rangle}_{\rho}^{\delta {\cal E}}} \simeq -(40 MeV)^3$, reducing ${{\langle
\bar q q \rangle}_{\rho}}$ to $- (203 MeV)^3$ from its vacuum value of $-(230
MeV)^3$. At this density, the kinetic and the interaction energy almost
cancel each other and there is saturation.  This is reflected on ${{\langle
\bar q q \rangle}_{\rho}^{\delta {\cal E}}}$ also. The stabilizing density
dependence of $g_\sigma ^\ast $ and $g_\omega ^\ast $ make $\mid {{\langle
\bar q q \rangle}_{\rho}^{\delta {\cal E}}} \mid $ smaller than that in the
Walecka model. Thus ${{\langle \bar q q\rangle}_{\rho}}/{{\langle \bar q q
\rangle}_0}$ reduces with density to be at par with the leading order as
shown in Fig.2 for ZM models. This is more pronounced for ZM3 Model where
${{\langle\bar q q \rangle}_{\rho}^{\delta {\cal E}}}$ increases slowly with
$\rho $ and the corresponding $\sigma _{eff}$ (Fig.3) is decreased.

If there is scaling as in eq.(\ref{eq:scaling3} or \ref{eq:scaling1}) or in
any other form, the effective nucleon mass $M^ \ast _N $ vanishes when the
condensate vanishes. This is awkward for nuclear physicists since the meaning
of nuclei or nuclear matter composed of zero mass particles is ill-defined.
Is there some model where the condensate will decouple ? 

Indeed, as shown in Fig. 1-2, the ZM models decouple {\it a decreasing
condensate} from the effective nucleon mass whereas in the Walecka model at
$\rho > \rho _0 $, $ {{\langle \bar q q \rangle}_{\rho}}/ {{\langle \bar q q
\rangle}_0}$ goes up and ${M^ \ast _N }/{M _N}\rightarrow 0$.  In-medium quark
condensate is governed by $\sigma _{eff}$ ( eq.(\ref {eq:fpimpi})) which in
turn is the derivative of $({\cal E}/ \rho)$ and not of $M_N^*$ (eq.(\ref
{eq:sigmaeff})). This explains why the ${\langle \bar q q \rangle}$ and
$M_N^*$ might have different density dependence and therefore the decoupling
occurs. After all, at $\rho\,=\,\rho_o $, when ${{\cal E}/ \rho_o - M_N} =
-15.75 MeV$, ${\langle \bar q q\rangle}$ is found to be the same for all
hadronic models with very different $ M_N^*(\rho_o)$.

To investigate the role of variation of meson-couplings with $m_q$ in the
behaviour of $ {\langle \bar q q \rangle}_{\rho}$, we have separated the
contributions of these variations in Fig. 4a and 4b respectively. It is
interesting that in the ZM3 model the two contributions almost cancel each
other upto $\rho = 4 \rho_o$ which means that in eq.(\ref{eq:qqrho}) the term
actually important is the first one
\begin{equation}
\frac{{m_\sigma}^2} {{g^*_\sigma}^2}(M_N - M^*_N) = \frac{{m_\sigma}^2}
{{g^*_\sigma}} \sigma
\label{eq:first}
\end{equation}
which is proportional to the sigma field. In this type of model the $\sigma$
field is related not only with the scalar density (as in Walecka and ZM
models) but have a new term depending on $\rho$, which comes from the new
$\sigma  - \omega$ interaction present in this model \cite{zm}. So we can
conclude that the condensate goes to zero in this model only because the
contributions coming from the $\rho _B^2$ are cancelled by the second term in
eq.(\ref{eq:qqrho}) which is proportional to
\begin{equation}
\frac{{m_\sigma}^2} {{g^*_\sigma}^2 M_N}(M_N - M^*_N)^2 = \frac{{m_\sigma}^2
\sigma ^2}{M_N}.
\label{eq:second}
\end{equation}
For the ZM model the variation of the scalar coupling with $m_q$ is essential
to get value of the condensate near the leading order behaviour.  However,
comparing the linear Walecka model with the non-linear models we use, we can
conlcude that the effective density dependent coupling constants present in
these models are mainly responsible for the condensate not going up at
intermediate densities ($\rho /\rho _o = 1 - 2.5$) as occurs in the Walecka
model.

In this context it is useful to mention that the density dependence of the
coupling constants may be traced back to Brueckner Hartree Fock approaches
\cite{suply}. In ZM models this dependence arises quite naturally from an
interesting theoretical scaling property. It is noteworthy that in the
original proposal of ZM3 \cite{zm}, the authors claim that the justification
for this model has to be understood in the spirit of chiral symmetry
restoration. The present work confirms this, the ${\langle \bar q q\rangle}$
being restored to zero at $\rho \sim \rho_o$.

The non-linear interaction terms, in a heuristic way incorporates the effect
of many-body forces leading to a suppression of the scalar field (Fig. 5). As
a result, the lower bent of the nucleon mass, $M_N$, is checked in the
medium. Moreover, in ZM3, even the vector field gets suppressed - holding
back the increase of the condensate.  Thus the {\it{condensate goes down at
high density}} whereas the nucleon mass does not change so much.  So, we have
a working model of relativistic nuclear matter which has the right behaviour
for ${\langle \bar q q\rangle}$ built into it.  Also it gives us the hope
that one may indeed construct more refined models with the same features
which combines the two desirable features of vanishing ${\langle \bar q
q\rangle}$ and non-vanishing effective nucleon mass.

\vspace{2cm}
{\bf 5. Chiral Symmetry Restoration and Nuclear Matter Equation of State}
\vspace{.5 cm} 

Let us analyse our results in more details.  From Fig. 2 we find that 
for the ZM3 model which is nearest to leading order, the $\sigma _{eff}$
changes only $10\%$ from $\sigma _N$ at the critical density $\rho _c \sim
3.8 \rho _0$. This means the contribution from interaction  $\delta \cal E$
must remain small even for high densities.  This can be seen clearly  from
the eq.(\ref{eq:sigmaeff}) when it is rewritten as

\begin{equation}
{\frac {d({\cal E}/ \rho)}{d M_N}}\,\,= 1 - {\frac {d({\delta \cal E}/ \rho)}{d
M_N}} \sim 1 - {\frac {({\delta \cal E}/ \rho)}{ M_N}}\,.
\label{eq:BEeff}
\end{equation}

Indeed, this is the case for ZM models as shown by the equation of state
(EOS, in short) given in \cite{zm}, where ${\cal E}/ \rho -M_N$ remains
closer to zero even at $\rho \sim 3-4 \rho _0$. In Walecka model, however,
this is not true.  $\cal E/ \rho$ starts rising very sharply at $\rho =
1.5 \rho _0$.  Obviously, the vector repulsion takes over the scalar attraction
in later.  In ZM3, where both fields are supressed at high density, it is no
wonder that the interaction part remains subdued.

It is noteworthy that the relation given by the eq. (\ref{eq:BEeff}) is
exact for $\rho = \rho _0$. This explains the existence of the model
independent point $\rho _0$ where ${\langle \bar q q \rangle}_\rho\,=\,0.69$
and $\sigma_{eff} \,=\,44.245$ MeV, keeping the binding energy per particle
fixed to -15.75 MeV and $M_N = 939 $ MeV.

>From the eq.(\ref{eq:BEeff}) we understand that $\sigma_{eff}$ and
${\langle \bar q q \rangle}_\rho$ are directly connected to the EOS.
Softer it is, the condensate moves closer to the leading order, realizing the
restoration of chiral symmetry at high density. There is one parameter which
governs the high density behaviour of the EOS for our models - that is the
incompressibility, $K= 9 dP/{d\rho }$ where P is the pressure.  Not only its
value at $\rho _0$ but also its variation with $\rho$ should be small to
obtain a softer equation of state. This is exactly what happens for ZM
models, where incompressibility is small ($K(\rho _0) = 156 MeV$ for ZM3 and
220 MeV for ZM) and there is a possibility of chiral symmetry restoration.
But this is impossible for Walecka model with its very stiff EOS. With these
observations we can indeed connect the behaviour of the quark condensate with
the nuclear matter parameters.

\vspace{.5 cm} 
{\bf 6. Results and Discussion for finite $T$}
\vspace{.5 cm} 
 
For finite temperature $T$, one must take into account the effect of
pion gas which dominates the finite $T$ vacuum \cite{gl}, \cite{dei}.
This is easily done in the models considered above using the factorization
hypothesis, assuming that the pion gas is non-interacting with the nucleons
at high temperature. At moderate $T$ this may introduce errors but at high
$T$, near the chiral restoration point surely this is not a bad
approximation. For moderate $T$ careful calculations have been done by
Leutwyler and Smilga \cite{ls} using the pion-nucleon scattering data through
dispersion relations and recently by  Koike \cite {ko} using QCD sum rules
including the effect of pion-nucleon scattering.  These studies show that the
nucleon mass does not change by more than a few percent at finite T, thus
supporting our effective model calculations.
 
In the present paper we restrict ourselves to the case of zero baryon
density and finite T. This is the situation where the vector meson does 
not couple to the nucleon-antinucleon soup, but the scalar density is
non-zero and an expression for the $\sigma _N(T)$ can be found easily.
Thermal pions in the vacuum can be effectively taken through the T-
dependence of $f_ \pi$ \cite{ko},
\begin{equation}
 {f_\pi }^2(T)={f_\pi }^2(1-\frac{T^2 B_{1}(\frac{m_{\pi}}{T}
)}{8{f_\pi }^2 })
\label{eq:fpiT}
\end{equation}
where 
\begin{equation} 
  B_{1}(z)\,=\,\frac{6}{\pi^2}\int_z^\infty \,dy
(y^2 - z^2)^{\frac{1}{2}}\frac{1}{e^{y} - 1}.
\label{eq:b1T}
\end{equation}

The definition of the $\sigma _N(T)$ for zero chemical potential is
as follows :
\begin{equation}
\frac{{\langle \bar q q\rangle}_T} {{\langle
\bar q q \rangle}_{\pi}} = 1 - \rho _s
\frac{\sigma _N(T)}{{m_\pi }^2{f_\pi }^2(T)}
\label{eq:qqT}
\end{equation}
and 
\begin{equation}
\rho _s \sigma _N(T) = [\frac{{m_\sigma}^2}
{{g^*_\sigma}^2} (M_N - M^*_N(T)) + (1 + \alpha )\frac{{m_\sigma}^2}
{{g^*_\sigma}^2 M}(M_N - M^*_N(T))^2] \sigma _N(T = 0).
\label{eq:sigmaT}
\end{equation}
where 
\begin{equation}
\rho _s =\frac{\gamma}{(2\pi)^{3}}\int
d^{3}k\,\frac{M^*_N(T)}{E^{\ast}(k)} (n_{k} + \bar n_{k}))
\end{equation}
is the scalar density.

Using $M^*_N(T) = M_N - g^*_{\sigma} \sigma(T)$ and $m^*=
{M^*_N(T)}/{M_N}$  in eq.(\ref{eq:sigmaT}) where the scalar field
$\sigma(T)\,=\,\, {\rho _s}\, m^{\ast^{\alpha}}{g^*_\sigma}/{{m_\sigma}^2} $
the sigma term reduces to a simple form :

\begin{equation}
\sigma _N(T) = (2+\alpha - (1+\alpha )\, m^*)\,m^{\ast^{\alpha}}\,\sigma
_N(0)  \,,
\label{eq:WZM}
\end{equation}

In Figs.6(a-c), the change of  quark condensate with T is shown, - with and
without the pion effects in the vacuum.  As in the case of finite density the
behaviour of the quark condensate  and the $M_N^*$ with temperature is found
to be different. The condensate goes to its zero value in all the three
models, - signalling perhaps, a new phase. The change in $M_N^*$ in ZM models
is very little compared to that in Walecka model. In the latter, the
condensate goes down very sharply, -just like the $ M^*_N(T) $.  The fall off
is not compensated by the rise in $\rho _s$ : non-linearity in ZM models
stabilizes $\sigma _N(T)$ to 50 MeV at critical temperature (Fig. 7) which
does not happen to linear Walecka model. This is also reflected in the 
$\sigma _N(T)$ vs. T plot (Fig. 8). For this model $\sigma _N(T)$ increases
sharply, meaning a sharp decrease in the condensate which indicates a first
order phase transition (Fig. 6a) at $T_c \sim 180 $ MeV.  The ZM models, on
the other hand, suggest a continuous and therefore a higher order transition.
The inclusion of pions in the vacuum reduces the critical temperature ($T_c$)
by $10 \%$. For example in ZM3 model, $T_c$ becomes 230 MeV from its pionless
vacuum-value $T_c = 260 MeV$.
 
\vspace{.5 cm} 
{\bf 7. Summary and Conclusion}
\vspace{.5 cm} 
 
In summary it is intriguing to find that the quark condensate at normal
$\rho$ is independent of the hadron-model chosen, - models with very
different $M_N^*$ and incompressibility. But at high $\rho$ and T it indeed
depends on the modelling and it is possible to get a decoupling of the
condensate from the effective nucleon mass. The fact that ${\langle \bar q q
\rangle}_\rho$ decreases with density in the ZM3 model (less than linearly)
shows that EOS must be soft favouring a smaller incompressibilty (less than
220 MeV). However, the decrease in ${\langle \bar q q \rangle}_\rho$ does not
compensate the linear rise of the density. Consequently, $\sigma _{eff}$ at
high $\rho$ reduces, by $10-20 \%$. The reduction at $\rho _0$ is rather small
in the hadron models we used. Further, pionic corrections arising from $\sigma$
and $\omega$ fields will be there- but as commented and observed in the
reference \cite {birse} - a calculation of $\sigma _{eff}$ with linear
sigma model - their net effect may be small.

With temperature, the condensate does go to zero. The $\sigma_N(T)$ changes
more dramatically only for the Walecka model which shows a sharper increase
compared to the non-linear hadron models where we have an increase of about
$10\%$ at $T=T_c$, suggesting the latter might have higher order phase
transition.


\noindent 
\vskip .5cm
{\bf Figure Captions}

Fig. 1 (a-b) : Scaling laws of eq.(\ref{eq:scaling3}) and eq.
(\ref{eq:scaling1}) respectively for Walecka  and Zimanyi-
Moszkowski models (ZM and ZM3).

Fig. 2 : Ratio of the condensates to the vacuum value as a function of
density for three models are shown.

Fig. 3 : $\sigma _{eff} $ with density. 

Fig. 4  : Leading order and other contributions to condensate ratios
of fig. 2 for (a) : ZM and (b) : ZM3 models.

Fig. 5  : The effective nucleon mass in the three nodels.

Fig. 6 : A plot of the ratio of the condensate with the vacuum value
and the effective mass with temperature for for (a) : Walecka, (b) ZM and (c)
: ZM3 models.

Fig. 7 : A plot of $\sigma _N$ with scalar density for three models.

Fig. 8 : A plot of $\sigma _N$ with temperature for three models.

\vskip 1cm

\noindent {\bf Table 1.}Coupling constants $C_{\sigma}^2$, $C_{\omega}^2$, 
$m^{\ast}(\rho _o)$ and ${{\langle \bar q q \rangle}_{\rho _o}}/ {{\langle
\bar q q \rangle}_0}$  for different models\\
with a fixed ${{\cal E}/ \rho_o - M_N} = -15.75$ MeV at $\rho _o =
0.15  fm^{-3}$.
\vskip .5cm
\begin{tabular}[pos]{|c|c|c|c|c|} \hline
\,\,\,\,\,\,models \,\,\,\,\,\,   &\,\,\,\,\,\, $C_\sigma^2$ \,\,\,\,\,\, & \,\,\,\,\
\,\, $C_\omega^2$ \,\,\,\,\,\, &  \,\,\,\,\,\, $ m^*(\rho_o)$\,\,\,\,\,\, &\,\,\,\,\,\,\
$ {{\langle \bar q q \rangle}_{\rho_o}}/ {{\langle \bar q q
\rangle}_0}$\,\,\,\,\,\,  \\ \hline
Walecka   & 352.69      & 269.70 & 0.54 & 0.69\\ \hline 
\,\,\,\,\,ZM  & 177.55 &  63.47 & 0.85 & 0.69\\ \hline
\,\,\,\,\,ZM3 & 440.43 & 303.22 & 0.72 & 0.69\\ \hline 
\end{tabular}\\
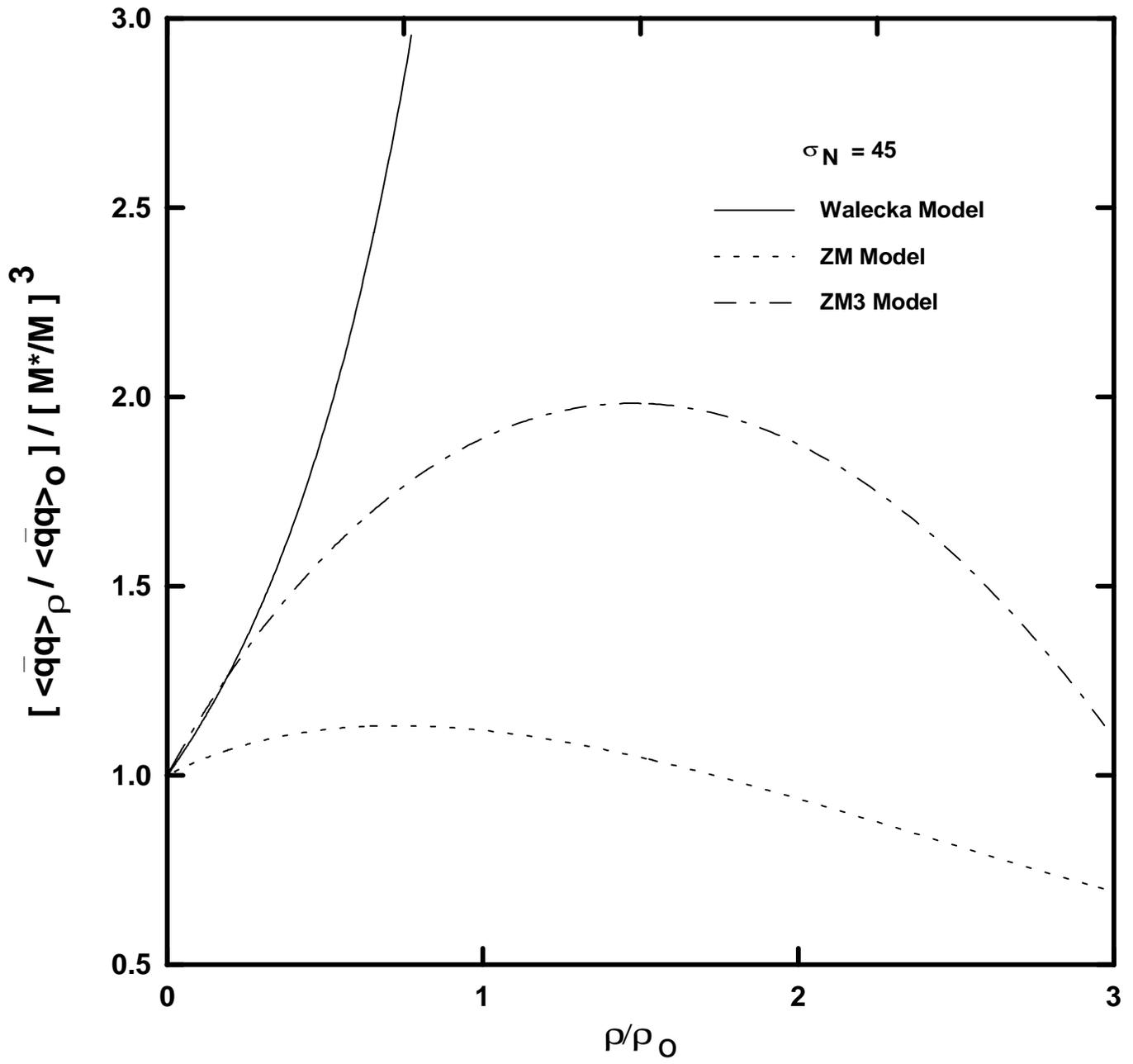
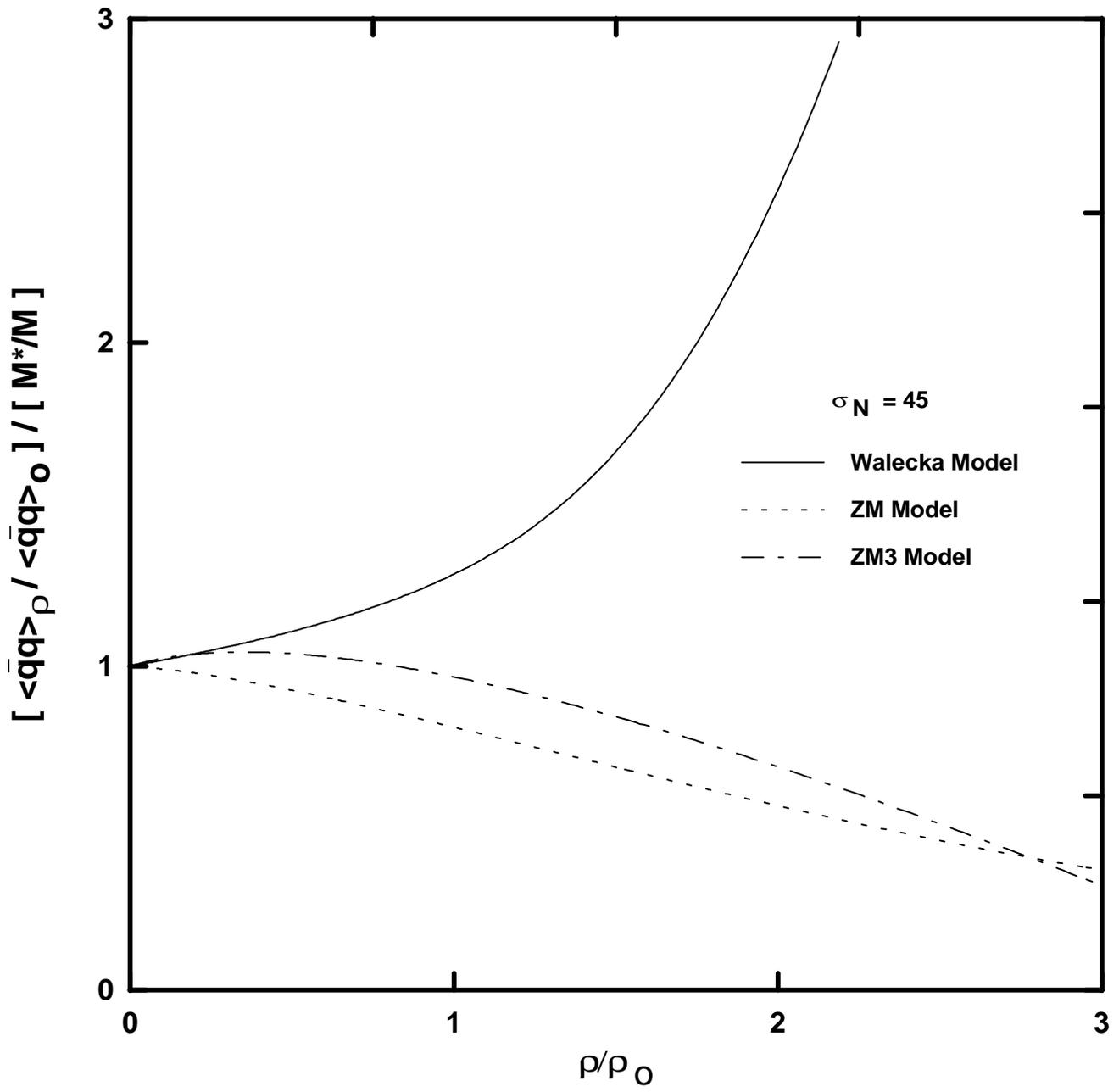
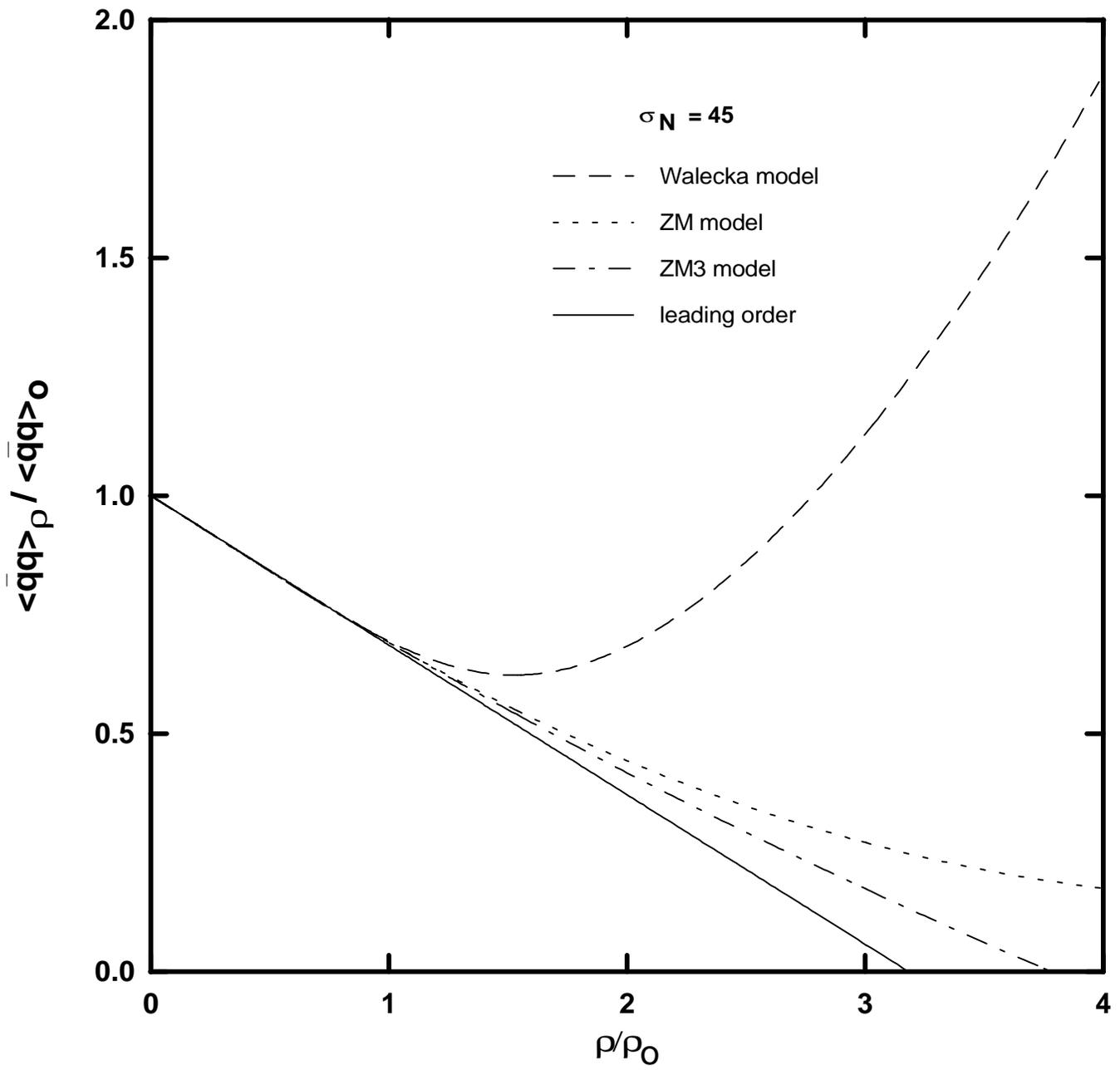
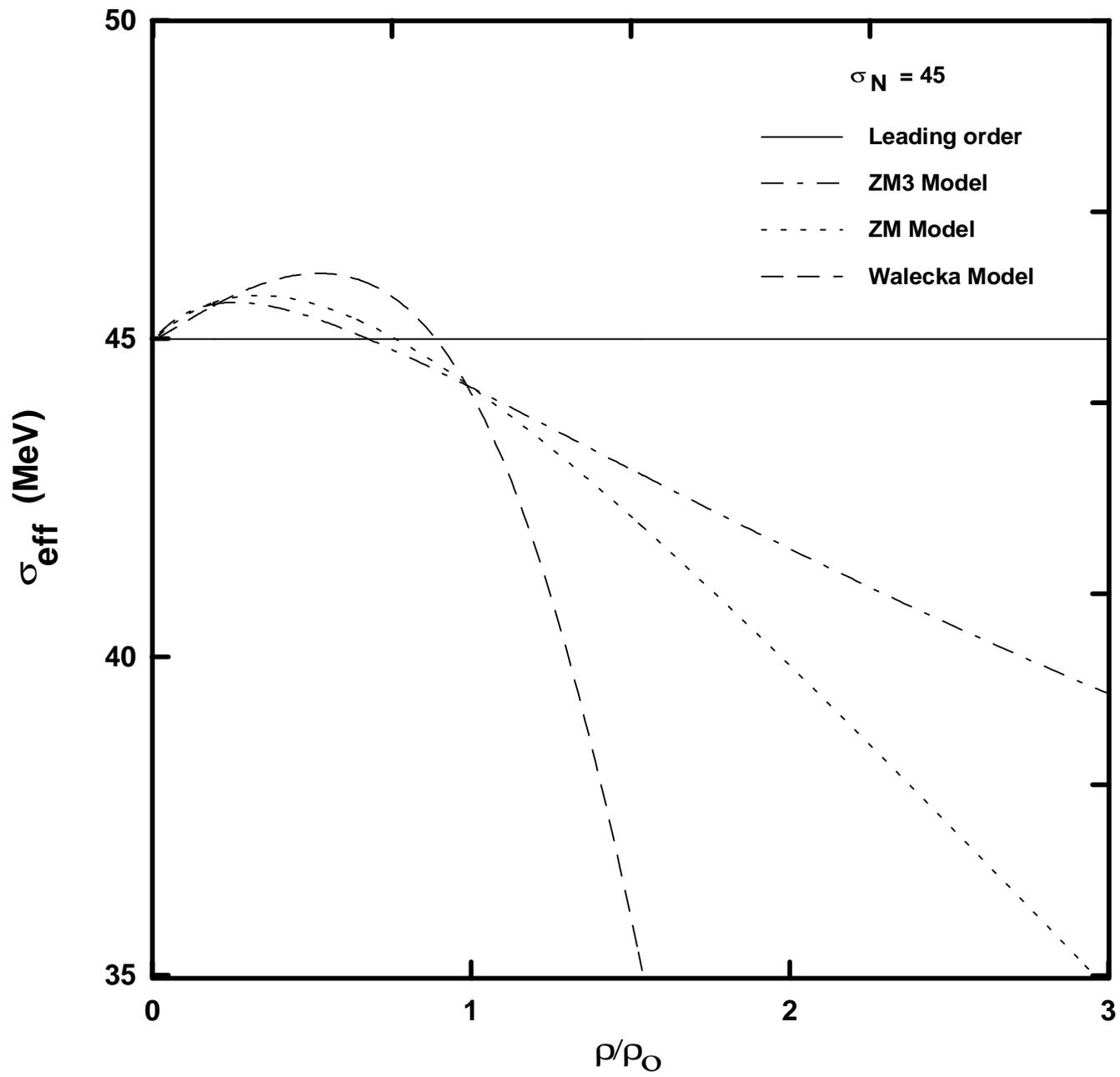
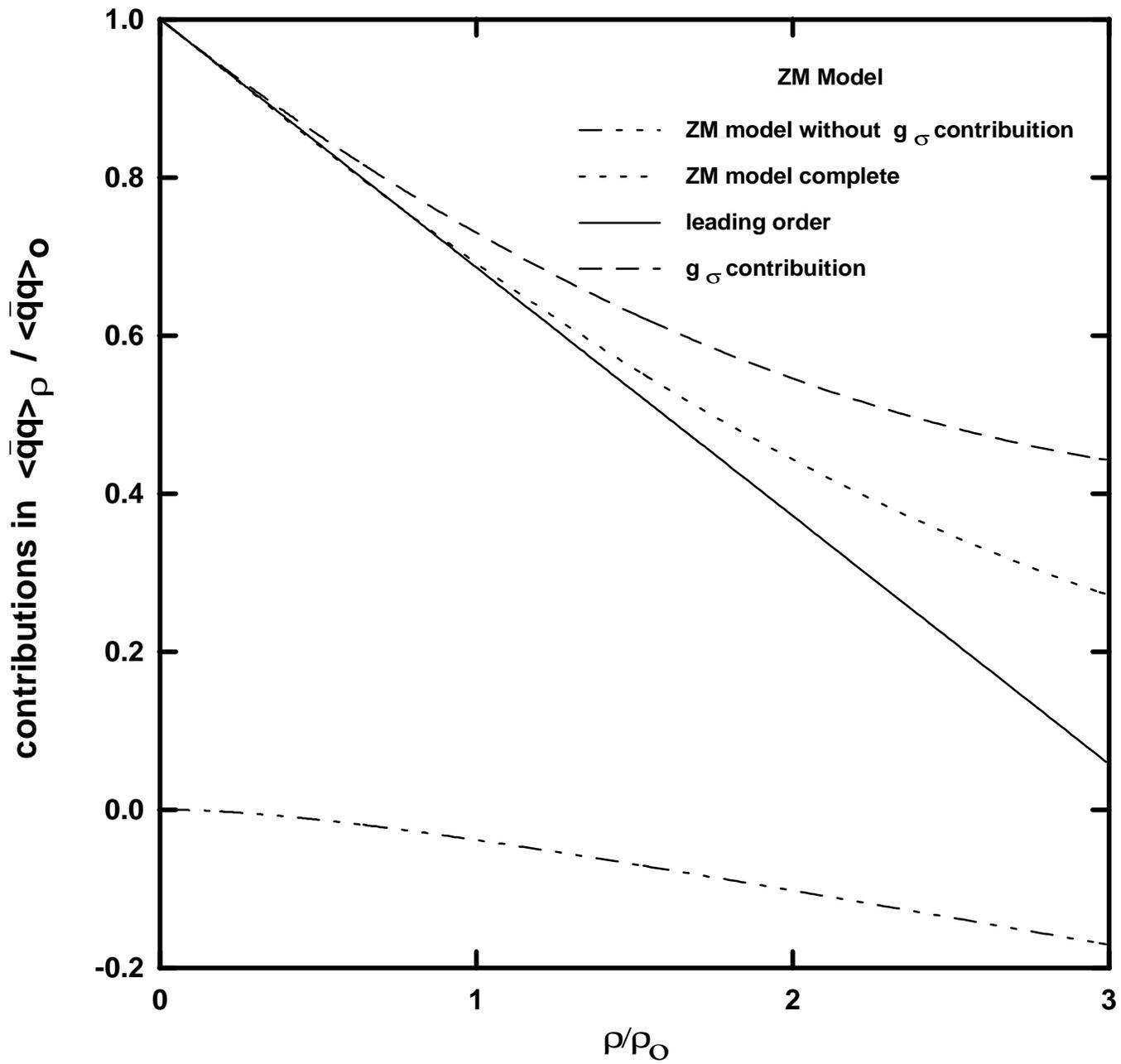
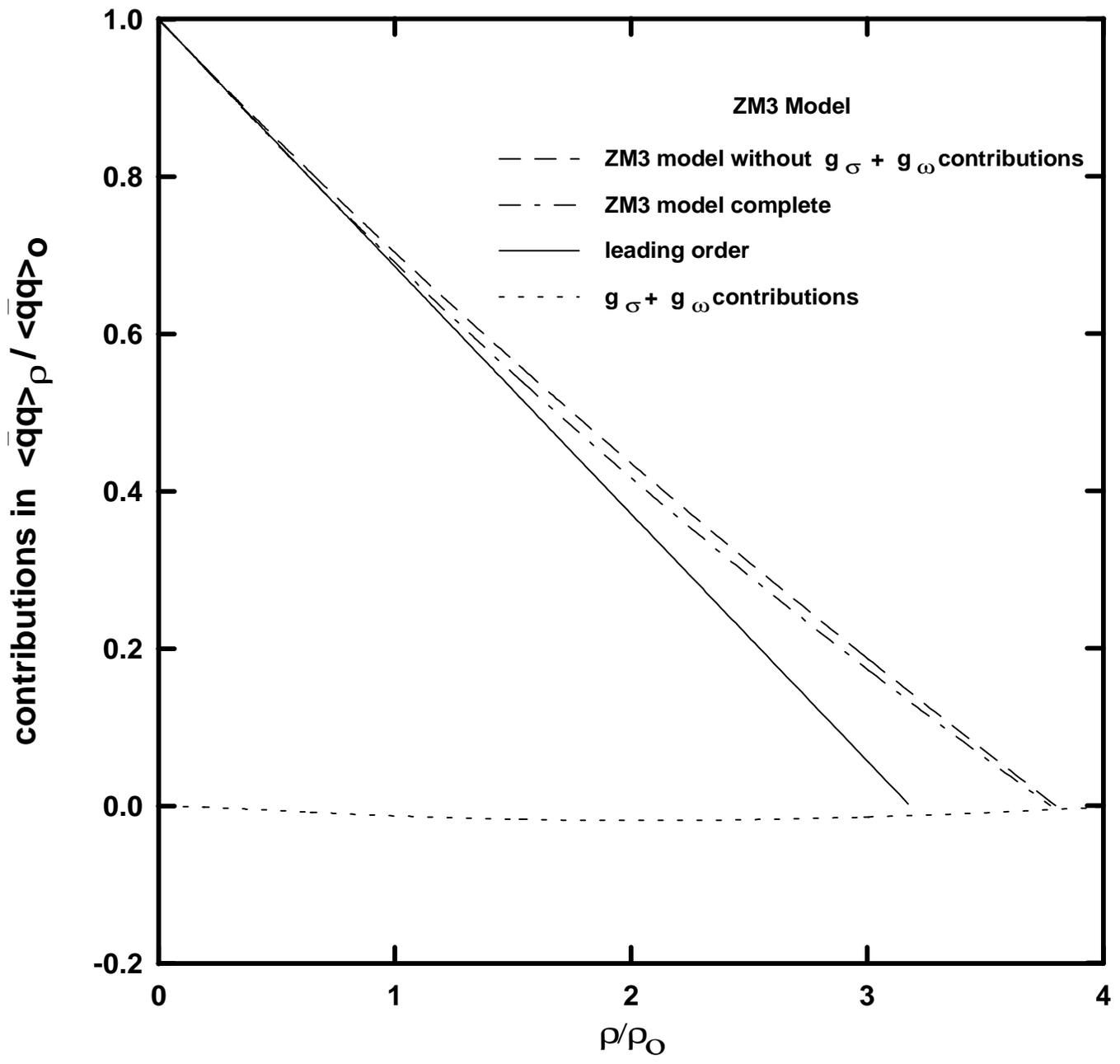
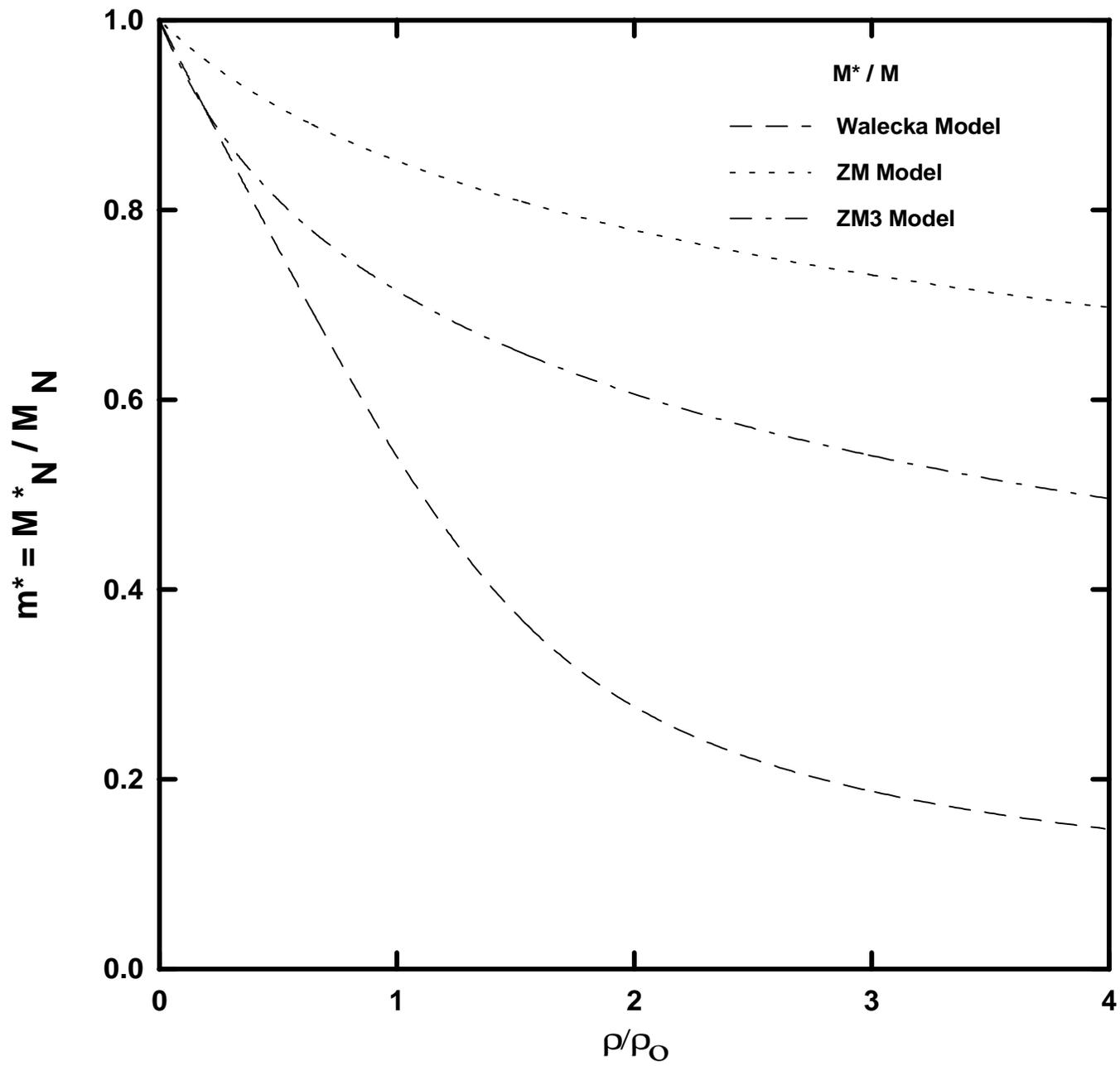
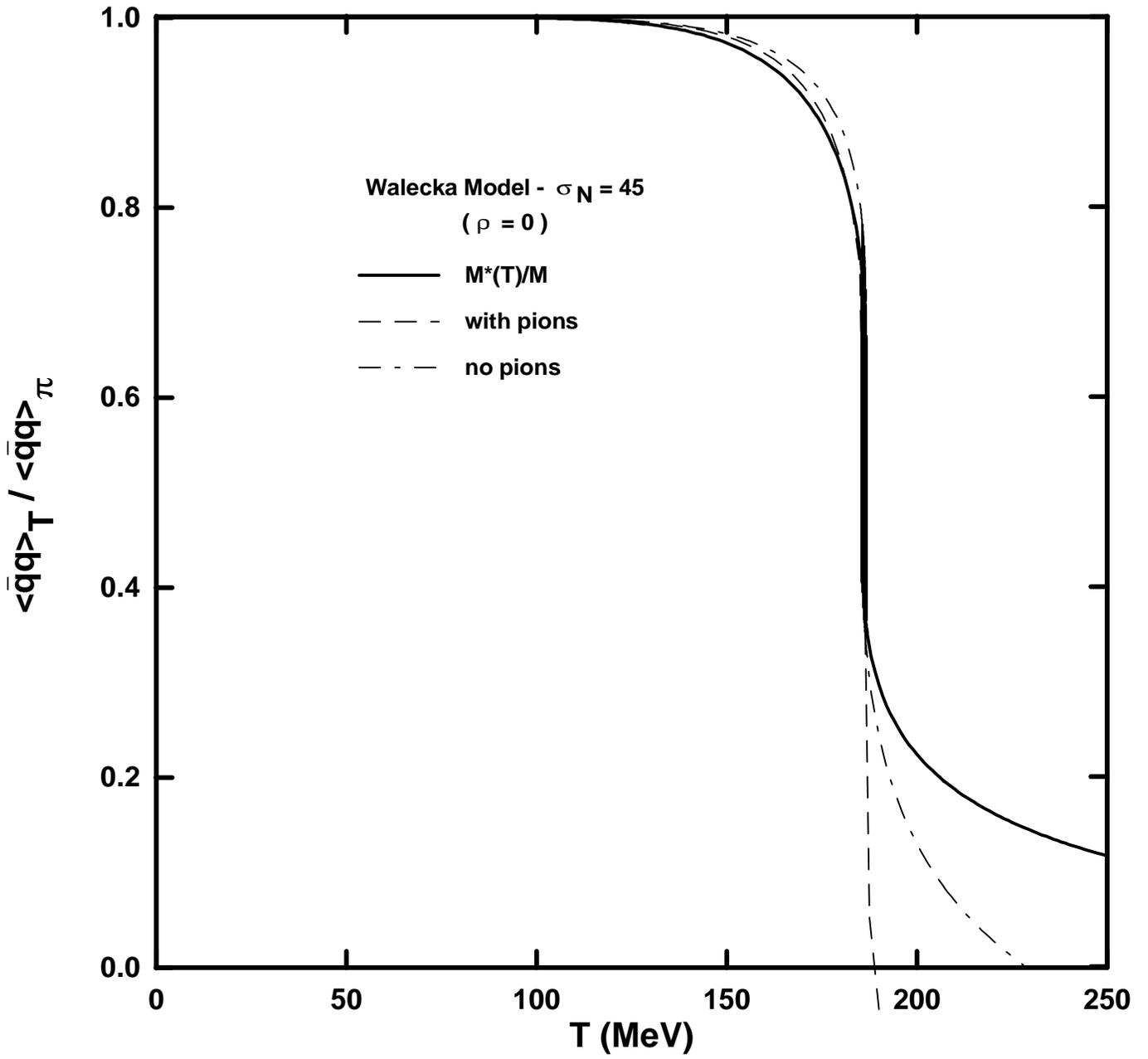
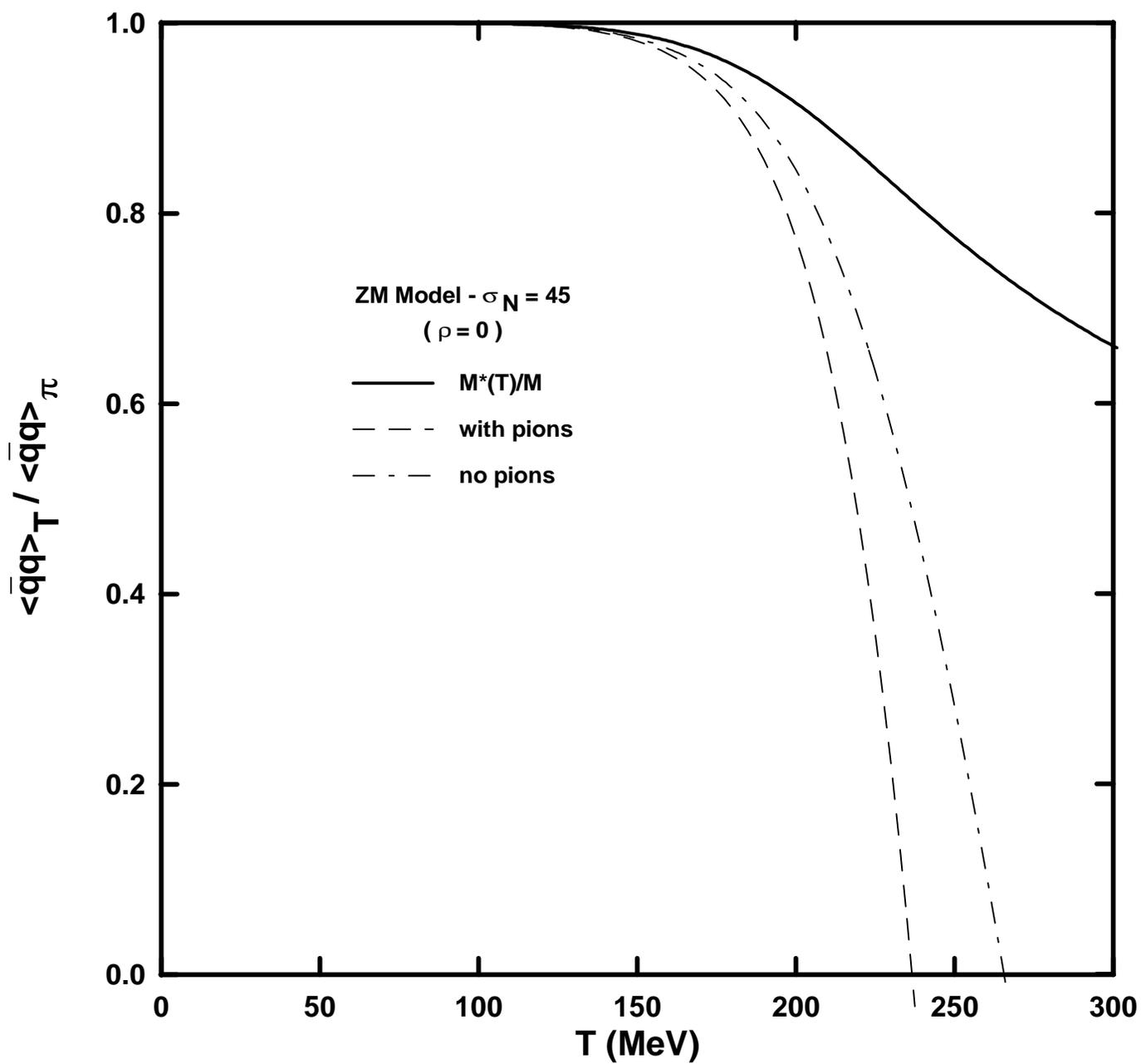
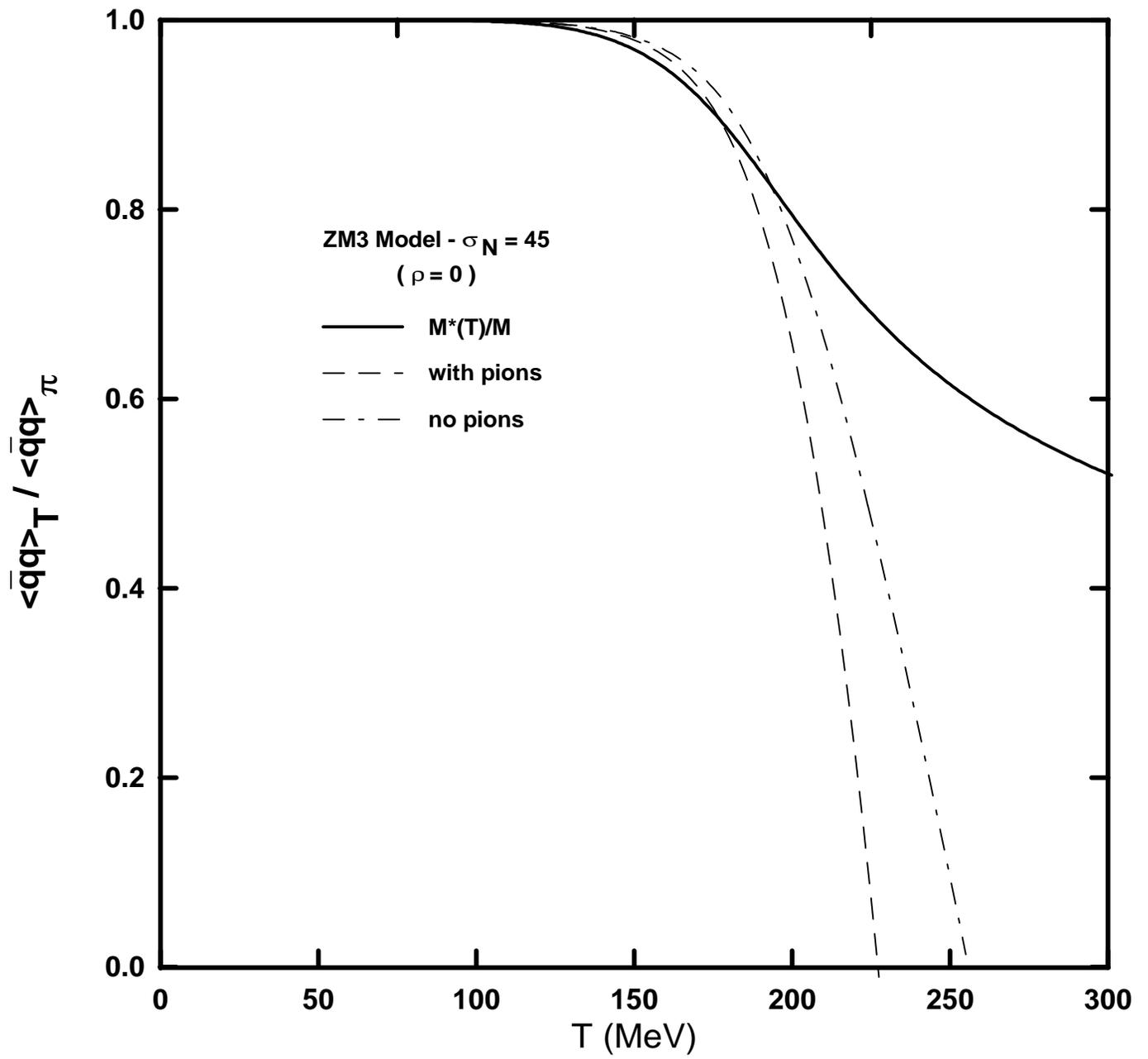
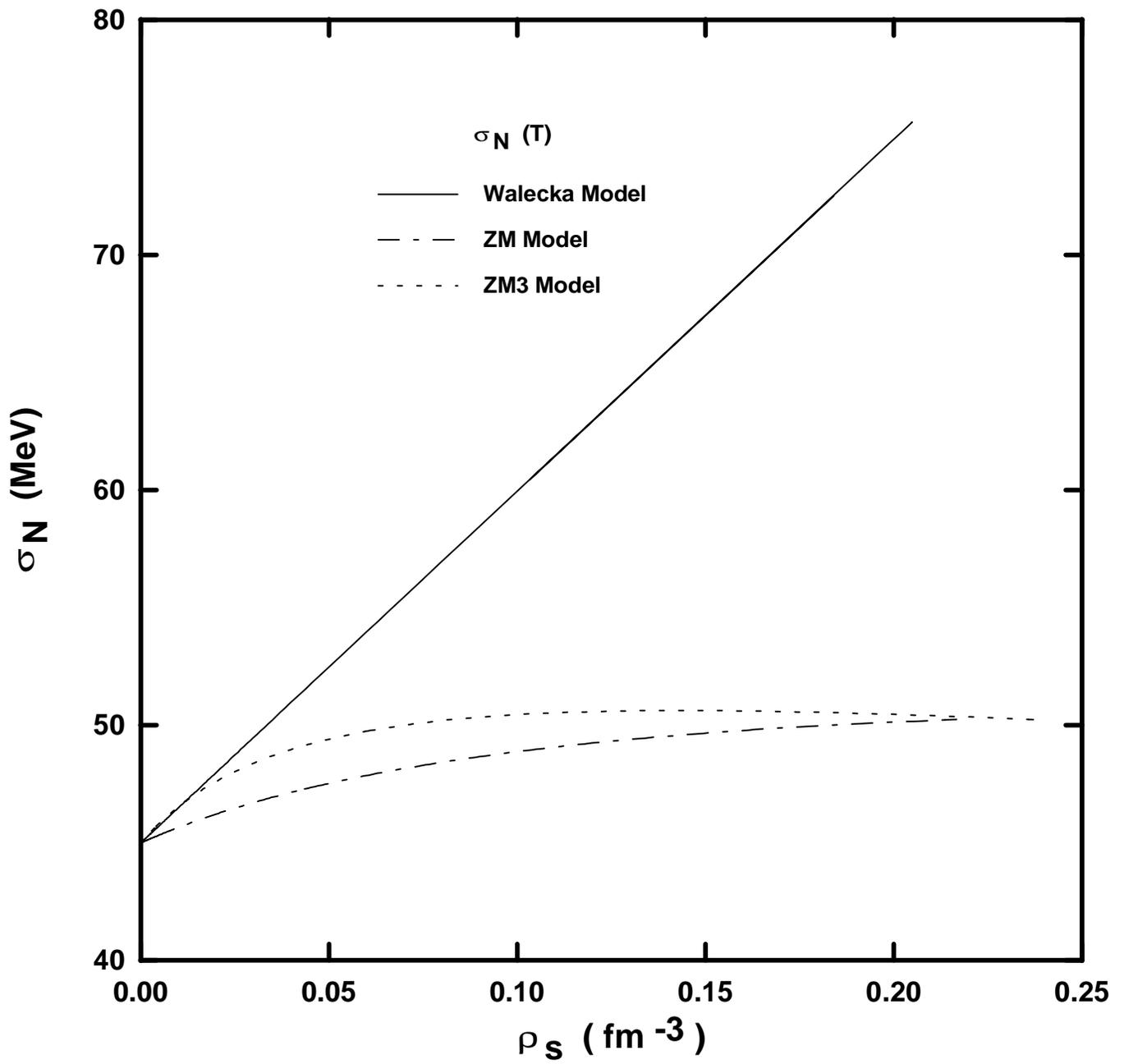
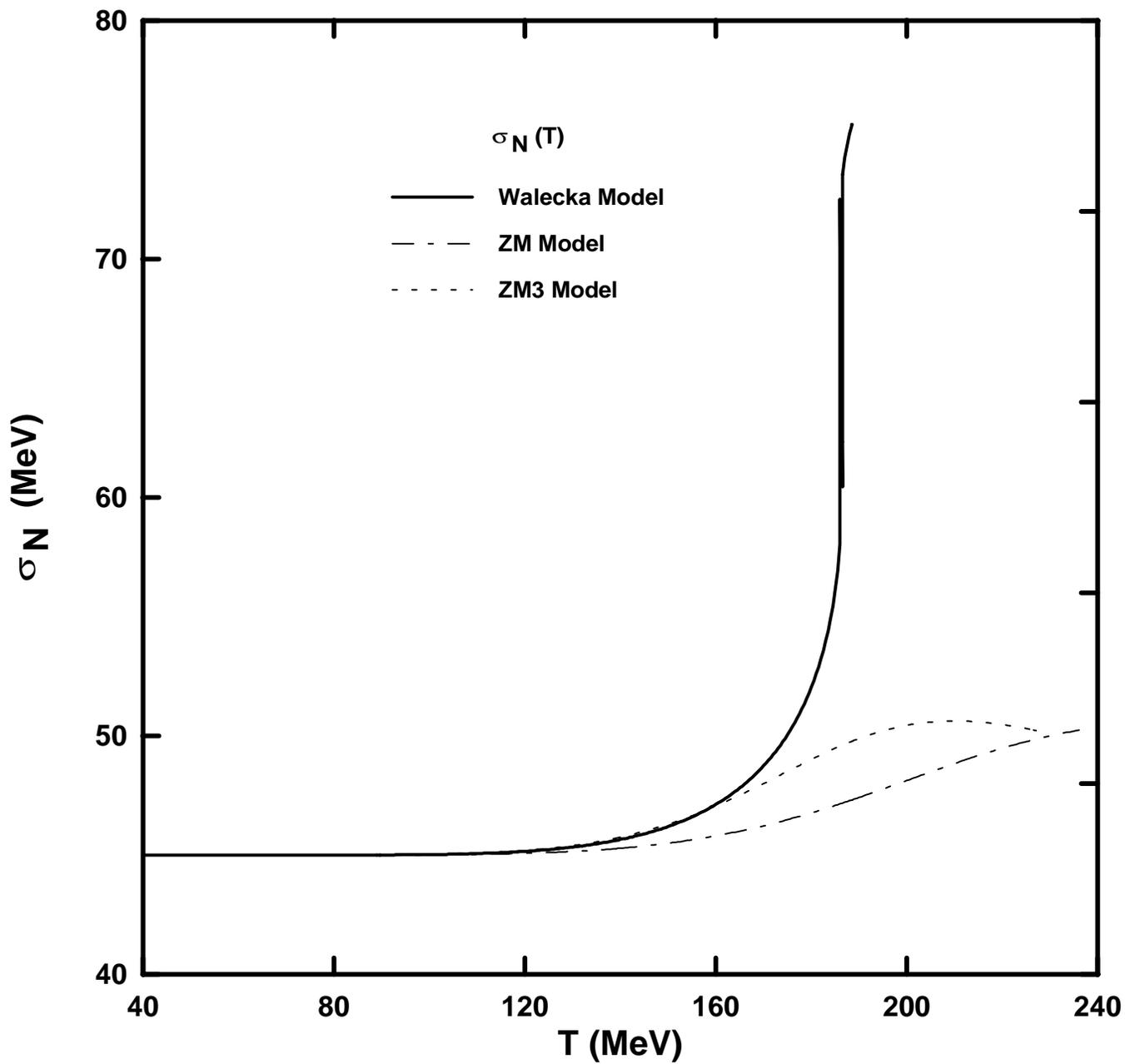
\end{document}